\documentclass[preprint]{aastex}

\begin{document}


\title{Atmospheric Turbulence Measurements with the Palomar Testbed Interferometer}


\author{R. P. Linfield}
\affil{Infrared Processing and Analysis Center, 100-22, California Institute of Technology,
    Pasadena, CA   91125}
\email{rpl@ipac.caltech.edu}

\author{M. M. Colavita}
\affil{Jet Propulsion Laboratory, 171-113, Pasadena, CA   91109}
\email{M.M.Colavita@jpl.nasa.gov}

\and

\author{B. F. Lane}
\affil{Department of Geological and Planetary Sciences, 150-21, California Institute of Technology,
Pasadena, CA   91125}
\email{ben@gps.caltech.edu}


\begin{abstract}
Data from the Palomar Testbed Interferometer, with a baseline length of 110~m and an observing wavelength
of $2.2\ \mu{\rm m}$, were used to derive information on atmospheric turbulence on 64 nights in 1999.
The measured two-aperture variance coherence times at $2.2\ \mu{\rm m}$ ranged from 25~msec to 415~msec 
(the lower value was set by instrumental
limitations---the interferometer could not operate when the coherence time was lower than this).  
On all nights, the spectrum of the short time scale 
($<600$~msec) delay fluctuations had a shallower spectrum than the theoretical Kolmogorov value of 5/3.
On most nights, the mean value of the power law slope was between 1.40 and 1.50.  Such a sub-Kolmogorov slope
will result in the seeing improving as the $\approx 0.4$ power of wavelength, rather than the slower $0.2$ power
predicted by Kolmogorov theory.

On four nights, the combination of delay and angle tracking measurements allowed a derivation of the (multiple) wind velocities 
of the turbulent layers, for a frozen-flow model.  The derived wind velocities were all $\le 4\ {\rm m\ s^{-1}}$, except for a 
small $10\ {\rm m\ s^{-1}}$
component on one night.

The combination of measured coherence time, turbulence spectral slope, and wind velocity for the turbulent layer(s) allowed
a robust solution for the outer scale size (beyond which the fluctuations do not increase).  On the four nights with 
angle tracking data, the outer scale varied from 6~to 54~m, with most values in the 10--25~m range.  Such small outer scale
values cause some components of visibility and astrometric errors to average down rapidly.

\end{abstract}



\section{Introduction}
Optical and near infrared interferometers, by virtue of their long baselines, can achieve angular resolutions better
than is possible with a single telescope.  However, the atmosphere imposes serious limits on interferometer
sensitivity, via the coherence length ($r_0$) and coherence time ($t_0$):  the length and time scale over which atmospheric
effects change significantly.  At optical and near infrared wavelengths, atmospheric phase variations are dominated
by the temperature/density fluctuations of the dry air component.

Conversely, interferometric observations can yield detailed information on atmospheric turbulence on spatial scales larger
than those of single apertures. In order to operate, interferometers need to measure (and correct) both delay and
angle fluctuations.  The time series of these measurements gives temporal and spatial information on atmospheric
turbulence.  The correlations in delay fluctuations over a finite baseline length can potentially 
provide additional information.

We used the Palomar Testbed Interferometer (PTI) \citep{col99}  for atmospheric measurements.  
PTI has three siderostats (only two of which
can be used at one time).  It can operate at wavelengths of~1.6 and~$2.2\ \mu{\rm m}$, with simultaneous operation possible.  
All the data reported in this paper
were taken on a 110~m baseline (oriented $20^\circ$ east of north), at a wavelength of $2.2\ \mu{\rm m}$
(2.0--2.4~$\mu{\rm m}$ passband).  Although the interferometer has a dual star mode (for narrow angle astrometry), only data
from single star mode was used in this analysis.

There were multiple motivations for this study.  The first was to better understand the physics of atmospheric refractivity
fluctuations.  The second was to quantify the atmospheric error in astronomical measurements (especially those with an
interferometer), and perhaps devise improved observational strategies.  The third motivation was to look for instrumental
error sources, by searching for deviations from an atmospheric signature.

\section{Observations and data selection}

\subsection  {Interferometer Delay Data}

During PTI operation, the position of the optical delay line, relative to a standard fiducial point, is monitored by laser
metrology, recorded at 500~Hz, and averaged in post processing to match the white-light sample time of 10~or 20~msec.
The delay line follows a predicted sidereal trajectory.  A correction is supplied by the 
output of the fringe tracker, which adjusts the
delay with a 5--10~Hz closed-loop bandwidth to follow the white-light fringe in the presence of atmospheric turbulence and
imperfect baseline and astrometric knowledge.

The fringe phase is measured for each 10~or 20~msec sample in a broadband, ``white-light'' channel, as well as in each
of~5 spectral channels across the band.  The white-light phase, which has a high SNR per sample, represents the error
of the fringe tracker; by adding the white light phase (scaled to delay units) to the measured delay line position, we 
obtain the total delay at 10~or 20~msec intervals.

The white-light phase measurements are all modulo-$2\pi$ radians; phase unwrapping maintains continuity during tracking, but
undetected unwrapping errors can occur due to low SNR or rapid atmospheric motion.  These are nominally detected and
corrected (one cycle at a time) at a lower rate, of order 1~Hz, using group delay measurement from the phases measured
across the 5~spectral channels \citep{col99}.  While the group delay provides an absolute measurement, the
low correction rate means that some fraction of the data can be off the central white-light fringe, with the fraction
growing for poor seeing.  We used four selection criteria to minimize such `cycle slips.'  First, sources with fringes weaker
than a threshold value were excluded.   Second,
only scans with an average of $<1.0$ detected fringe lock breaks per 24~s output record were used.  This criterion biased our
results, by excluding approximately one third of the total data: 
nights (and segments of nights) with noisy atmospheric conditions.  However, it was necessary in order
to avoid contamination from instrumental effects.  Third, whenever group delay measurements indicated a cycle slip in the
fringe tracking, 0.5~s of data at that epoch were flagged and not used in the analysis.  Fourth, sources brighter than a 
threshold value were not used.  Fringe data is recorded when locked on the white-light fringe, as ascertained by the
fringe signal-to-noise ratio.  Bright sources could be tracked on sidelobes of the central fringe with the current
implementation of fringe centering.

\subsection  {Star Tracker Angle Data}

During interferometer operations, the angle of each starlight beam was sensed (relative to the optical axis on the beam
combiner table) every 10~msec.  This ``error signal'' was used to drive a fast steering mirror, with a closed loop
bandwidth of 5~Hz.
The fast steering mirror (FSM) position was desaturated into the siderostat positions with a bandwidth of
$\approx 0.1$~Hz.

The three component signal complicated the use of the star tracker data for atmospheric measurements.  Fortunately,
the atmospheric timescales of interest were shorter than the timescale for desaturation of the FSMs.  It was not feasible
to add the error signal to the FSM positions.  With the quad-cell sensor used for angle tracking, the gain is a function
of the seeing-dependent spot size.  While not an issue for a conservative closed-loop system, this does prevent accurate
combination of the error with the mirror position to get angle variations at all frequencies.
Therefore, we used the FSM positions by themselves, with separate analysis of the error signals in order to set
limits on short timescale angle variations.

We used star tracker data only during periods when fringes were being tracked ({\it i.e.} the time gaps in the fringe white light
files were used as a flag to edit out similar time periods in the star tracker data).

\section {Data Analysis}

\subsection {Delay Data}

\subsubsection  {Sidereal Fit}

The time series of delay (and phase) measurements contained large (tens of meters) geometric components, in addition to  
atmospheric components.  To remove the geometric component, we subtracted a least squares sidereal fit from all the data on 
a source on a given night:
\begin{displaymath}
a + b\sin({\rm ST}) + c\cos({\rm ST})
\end{displaymath}
Here, ST represents sidereal time.  By solving for three parameters ($a$, $b$, and $c$), our residuals were insensitive to
uncertainties in the length or orientation of the interferometer baseline, or to the zero point in the delay line metrology.

In order to avoid removing significant short term ($<100$~s) atmospheric signature from our data in the fitting process,
we only used sources with 
multiple scans on the same night.  The time span for the sidereal fit was therefore always $>1000$~s.  Numerical tests
showed that a sidereal fit over a duration $T_{\rm fit}$ caused a noticeable suppression in the structure function of the
residuals on time scales as short as $T_{\rm fit}/10$, with larger effects on longer timescales.

\subsubsection {Structure Functions}

Using the residual ({\it i.e.} post sidereal fit) delay/phase time series, structure functions $D_\tau(\Delta t)$ were
calculated:
\begin{equation}
D_\tau(\Delta t)\equiv \left<\left[\tau(t+\Delta t)-\tau(t)\right]^2\right>
\end{equation}
Here $\tau(t)$ is the residual delay at time $t$, and the $<>$ brackets denote ensemble averaging.  With observations
scheduled for amplitude visibility measurements (the majority of our data), the scan lengths were typically 130~s.
One structure function was calculated for each scan.  On some nights, long scans (20--30~min.) were made, solely for
atmospheric measurements.  For those scans, structure functions were calculated for each 3~minute segment of data.

\begin{figure}
\figurenum{1}
\plotone{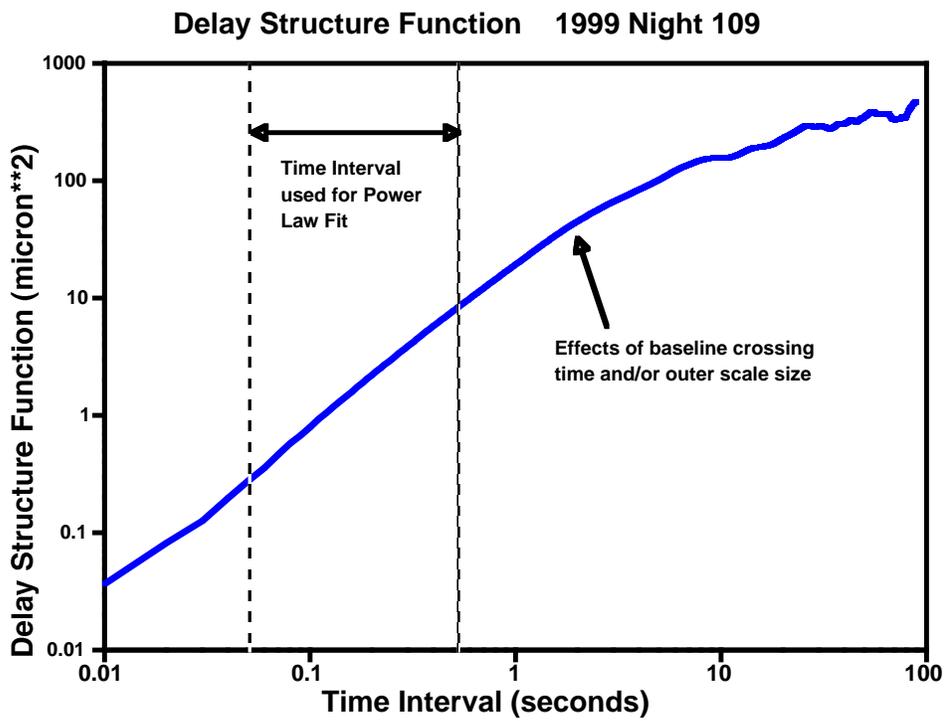}
\caption{Typical delay structure function.  The fitted parameters for this scan are a slope of 1.46 and a coherence
time of 122 msec. \label{fig1}}
\end{figure}

A typical delay structure function is shown in Figure~\ref{fig1}.  On timescales from 50~msec to $\approx 1$~s, a 
clean power law slope was seen in the structure
function of nearly every ($>90$\%) scan.  On the shortest timescales ($<50$~msec), the structure function for many scans 
exhibited power above that extrapolated from the slope at longer (50--500~msec) timescales.  We interpret this
excess power as being due to instrumental effects ({\it e.g} vibrations).  
On timescales longer than $\approx 1$~s, the slope of $D_\tau$ decreased,
due to some combination of outer scale length and
baseline crossing effects ({\it i.e.} the product of wind speed and time interval becomes comparable to or greater than
the baseline length).  The long baseline (110~m) of PTI, compared to that of other optical/IR interferometers, has a longer
wind speed crossing time.  There is therefore a 
relatively long time interval ($>1$~s) over which the fluctuations at the two siderostats are uncorrelated.  As discussed
later, outer scale effects were more important than the effect of a finite baseline length, at least
for the four nights on which we recorded star tracker data from both siderostats, and were able to measure the outer
scale length.

An alternate method of quantifying fluctuation statistics is power spectral density.  Structure functions have the advantage
of being unaffected by gaps in the data, and allow a more direct calculation of the dependence of seeing on wavelength, or
astrometric precision on baseline length and integration time.  The comparison between power spectral density and structure
functions is discussed in section~\ref{outerscale}.

\subsubsection {Extracting slope and coherence time}

A least squares fit to the slope of $D_\tau$ was made for each scan, over the interval 50--500~msec (10~msec integration mode)
or 60--600~msec (20~msec integration mode).  Equal weight was given to each logarithmic time interval in the fit.  Results from
a given scan were not used if the rms residual to the fit was $>0.02$ in log-log space, or there was too little data
(time span $<100$~s or $>40$\% of the data from the span missing).  Scans with large residuals did not exhibit a simple
power law turbulence spectru.   As a result, these scans could not be accurately characterized by a single power law index, and we 
chose not to use them in further analysis.

The two parameters from the fit were the slope and intercept.  We wished to derive the coherence time $T_{0,2}$, 
defined as the time
interval over which the interferometer phase fluctuations have a variance of $1\ {\rm radian}^2$.  Our notation follows that
in \citet{col99}; the ``2'' refers to the contributions from the two apertures of the interferometer.

The variance $\sigma^2_\tau(T)$ of delay over a time interval $T$ is \citep{tre87}:
\begin{equation}
\sigma^2_\tau(T)={1\over T^2}\int_0^T (T-\tau)D_\tau(t)dt
\end{equation}
If $D_\tau(\Delta t)=c_0(\Delta t)^\beta$,
\begin{equation}
\sigma^2_\tau(T)={c_0T^\beta\over (1+\beta)(2+\beta)}
\end{equation}
If $D_\tau(t)$ is expressed in ${\rm radian}^2$, we obtain
\begin{equation}
T_{0,2}=\left[{(1+\beta)(2+\beta)\over c_0}\right]^{1/\beta}
\end{equation}
We can convert from our measured two aperture variance coherence times $(T_{0,2})$ at $2.2\ \mu{\rm m}$ to one aperture
difference times $(\tau_{0,1})$ at $0.55\ \mu{\rm m}$, as measured in adaptive-optics applications.  Setting
$c_0\tau_{0,1}^\beta=2(0.55/2.2)^2$, we get
\begin{equation}
\tau_{0,1\ (0.55\ \mu{\rm m})}=\left[{0.125\over(1+\beta)(2+\beta)}\right]^{1/\beta}T_{0,2\ (2.2\ \mu{\rm m})}
\end{equation}

Our data were taken over a range of zenith angles (ZA), up to ${\rm ZA}\approx 40^\circ$.  The value of $D_\tau$ 
at short time scales for a vertical
column of uniform turbulence is expected (on theoretical grounds) to vary 
linearly with the thickness of the column, suggesting that $D_\tau$ will be proportional
to $\sec({\rm ZA})$.  However, if the wind velocity is not perpendicular to the source azimuth, the dependence will be more
gradual.

The extreme range of the variation in $T_{0,2}$ from zenith angle variations will be (see eq.[4]) a factor of 
$(\sec 40^\circ)^{1/\beta}$.
For the values of $\beta$ measured at PTI ($\beta\approx 1.45$), this will be $\approx 1.2$.  As shown in 
section~\ref{results}, the range 
of coherence time was much larger than this, even within one night.  Therefore, only a small part of the observed variation
in $T_{0,2}$ can be due to the range in observed sky directions.  The fitted slope should not depend on zenith angle, based on
theoretical grounds.  We see no dependence in our data.

\subsubsection {Contributions from internal seeing}

Light from the siderostats is brought into the optics lab via evacuated beam tubes.  Within the optics lab, the delay lines
operate in air.  A ``building within a building'' design \citep{col99} helps to keep this air still and nearly
isothermal.  A test of the fluctuations within the optics lab, including the long delay lines, gave a delay structure function
1000~times smaller than those on natural stars (M. Swain, private communication).  We conclude that
refractivity fluctuations within our instrument gave a negligible contribution to our derived atmospheric parameters.

\subsection  {Star Tracker Data}

\subsubsection {Structure Function of Angle Data}

\begin{figure}
\figurenum{2}
\plotone{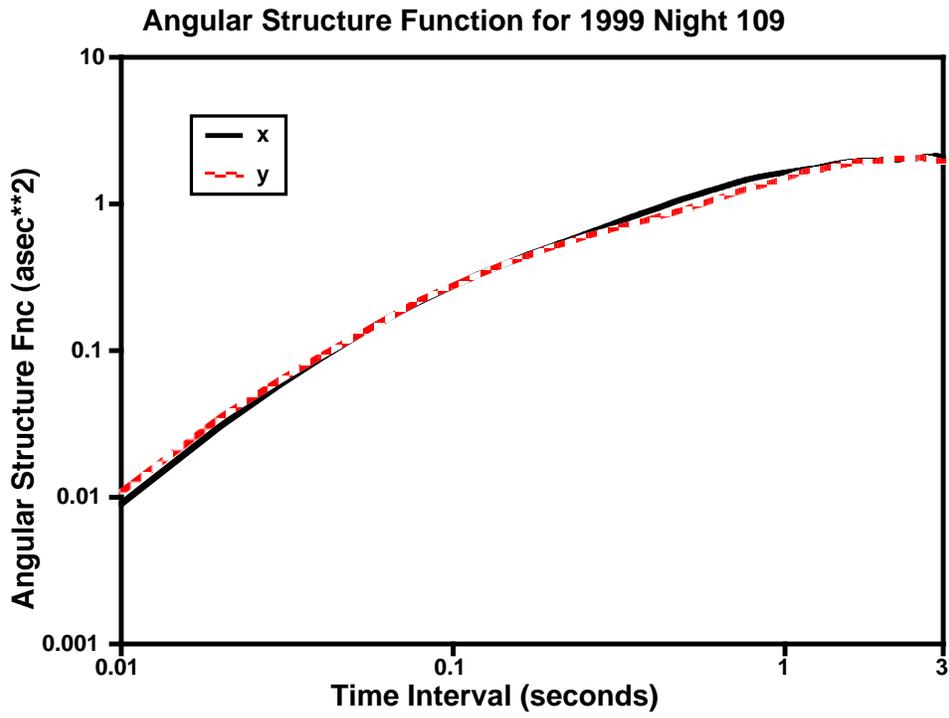}
\caption{Typical measured angular structure function.  Data from the North star tracker fast steering mirrors
are plotted. \label{fig2}}
\end{figure}

The time series of starlight arrival angle measurements $\alpha_x(t)$ and $\alpha_y(t)$, in directions $x$ and $y$, 
were used to derive angular structure functions,
$D_{\alpha_x}\left(\Delta t\right)$ and $D_{\alpha_y}(\Delta t)$:
\begin{equation}
D_{\alpha_x}(\Delta t)\equiv\left<\left[\alpha_x(t+\Delta t)-\alpha_x(t)\right]^2\right>
\end{equation}
Separate structure functions were calculated for the Fast Steering Mirror positions and for the error ({\it i.e.} residual)
values. Typical FSM angular structure functions are shown in Figure~\ref{fig2}.  
$D_{\alpha_x}(\Delta t)$ and  
$D_{\alpha_y}(\Delta t)$ rise rapidly and then flatten
out.  The upturn at the longest time scales ($>0.3$~s) is due to the desaturation of the FSMs into the siderostats.

The structure functions for the star tracker error signals showed a sharp rise up to a constant ``plateau'' for all the scans.
This plateau was reached at a time scale of 20--50~msec, depending on the scan.  On time scales longer than $\approx 20$~msec
(full range 18--25~msec), the structure functions of the FSM positions were larger than the structure functions of the error
signals.  Therefore, the FSM angular structure functions ({\it e.g.} Figure~\ref{fig2}) 
represent all of the actual angle fluctuations,
except for a small amplitude, rapid component.

\subsubsection {Modeling $D_\alpha$}

We modeled the measured angles as a least squares fit for the wavefront slope across our apertures, in two ($x$ and $y$)
directions \citep{sar90}.  For the $x$-axis slope, we want to minimize
\begin{displaymath}
S\equiv\int_0^{2\pi}d\phi\int_0^R r\left[\tau(r,\phi,t)-c_0-x\alpha_x(t)\right]^2dr
\end{displaymath}
Here $R$ is the radius (20~cm) of our aperture, and $\tau(r,\phi,t)$ is the delay at position $(r,\phi)$ and time $t$ on the
aperture.  Setting $\partial S/\partial\alpha_x(t)=0$, we obtain
\begin{equation}
\alpha_x(t)={4\over \pi R^4}\int_0^{2\pi}\cos\phi d\phi\int_0^R r^2\tau(r,\phi,t)dr
\end{equation}
The definition of $D_{\alpha_x}(\Delta t)$ gives
\begin{equation}
D_{\alpha_x}(\Delta t)=2\left<\alpha_x^2(t)\right>-2\left<\alpha_x(t+\Delta t)\alpha_x(t)\right>
\end{equation}
Equation(7) leads to:
\begin{equation}
\left<\alpha_x^2(t)\right>={16\over\pi^2R^8}\int_0^{2\pi}\cos\phi d\phi\int_0^{2\pi}\cos\phi' d\phi'
\int_0^R r^2 dr\int_0^R r'^2\left<\tau(r,\phi,t)\tau(r',\phi',t)\right>dr'
\end{equation}
The analog to equation(8) for the spatial delay structure function $D_\tau(\Delta\vec x)$ is:
\begin{equation}
D_\tau(\Delta\vec x)=2\left<\tau^2(\vec x)\right>-2\left<\tau(\vec x+\Delta\vec x)\tau(\vec x)\right>
\end{equation}
$\left<\tau^2(\vec x,t)\right>$ is assumed to be independent of $\vec x$ and $t$, so that this term integrates to zero.
\begin{equation}
\left<\alpha_x^2(t)\right>=-{8\over\pi^2R^8}\int_0^{2\pi}\cos\phi d\phi\int_0^{2\pi}\cos\phi' d\phi'
\int_0^R r^2 dr\int_0^R r'^2 D_\tau(|(r,\phi,t)-(r',\phi',t)|)dr'
\end{equation}
With a similar derivation for $\left<\alpha_x(t+\Delta t)\alpha_x(t)\right>$, we get
\begin{equation}
D_{\alpha_x}(\Delta t)={16\over \pi^2R^8}\int_0^{2\pi}\cos\phi d\phi\int_0^{2\pi}\cos\phi' d\phi'\int_0^R r^2 dr
\int_0^R r'^2\left[D_\tau(B)-D_\tau(A)\right]dr'
\end{equation}
\begin{eqnarray}
A\equiv & |(r,\phi,t)-(r',\phi',t)| \nonumber \\
B\equiv & |(r,\phi,t+\Delta t)-(r',\phi',t)| \nonumber
\end{eqnarray}
We relate the spatial and temporal structure of the turbulence field with the frozen flow (Taylor) approximation.
For sky directions near zenith (as in our observations), the dependence of model $D_{\alpha_x}(\Delta t)$ and 
$D_{\alpha_y}(\Delta t)$ values on wind azimuth is relatively minor:  $\pm 20$\% in amplitude and the timescale of the
slope break, compared to the values for a $45^\circ$ azimuth.  
We therefore used an azimuth of $45^\circ$ in the calculations.  The close agreement between the measured
$D_{\alpha_x}(\Delta t)$ and $D_{\alpha_y}(\Delta t)$ curves suggests that the true wind azimuth was not near
$0^\circ$ or $90^\circ$ (or, more likely, that there was a mix of wind azimuths in the turbulent region of the atmosphere).

\subsubsection {Fitting for the wind velocity of the turbulent layers}

\begin{figure}
\figurenum{3}
\plotone{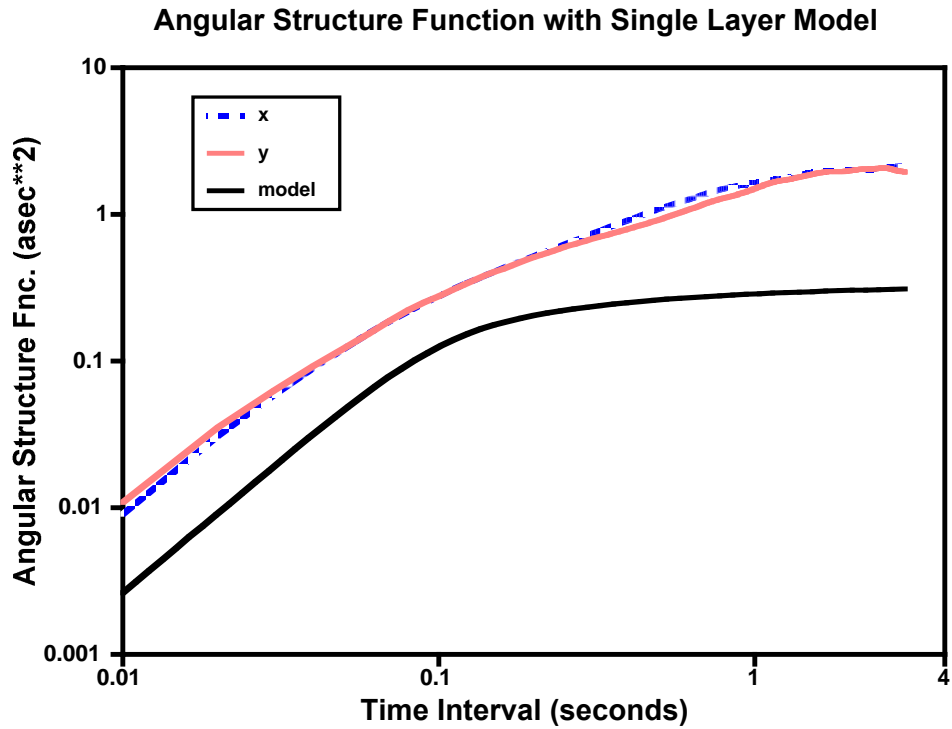}
\caption{Angular structure function from 1999 Night 109 with a single layer ($2.5\ {\rm m\ s^{-1}}$ wind velocity) model.  The
fit is poor on time scales $>0.2$~s.  The overall scaling mismatch is consistent with the amplitude calibration uncertainty
of the star tracker. \label{fig3}}
\end{figure}

\begin{figure}
\figurenum{4}
\plotone{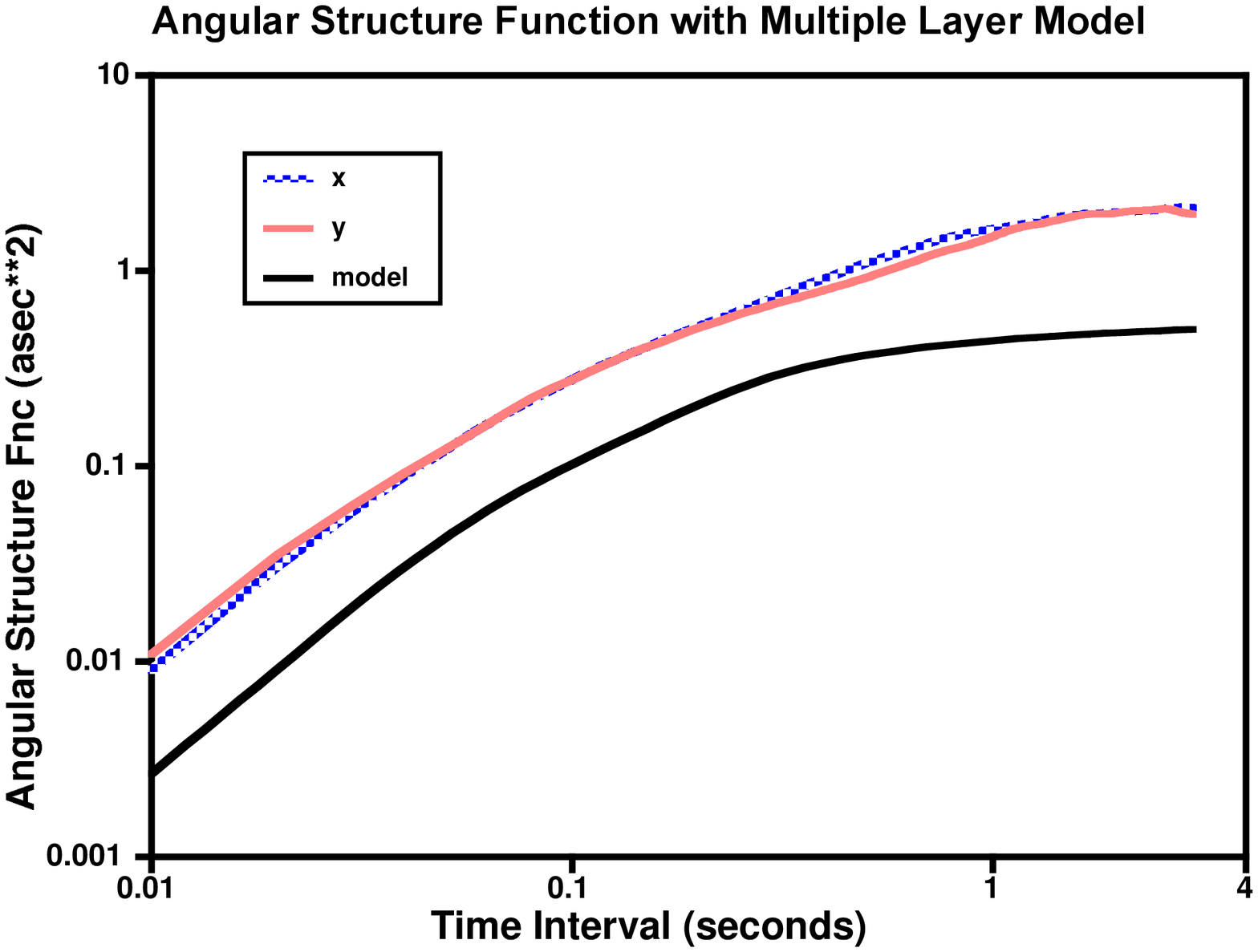}
\caption{Angular structure function from the same scan as in Figure~3, with a multiple layer 
(1 and $4\ {\rm m\ s^{-1}}$ wind
velocities) model.  The discrepancy between the data and the model on time scales $>0.5$~s is due to desaturation of the
fast steering mirrors into the siderostats.  The overall scaling mismatch is consistent with the amplitude calibration 
uncertainty of the star trackers. \label{fig4}}
\end{figure}

Once $D_\tau(\Delta t)$ is specified, the only free parameter in modeling $D_{\alpha_x}(\Delta t)$ is the wind velocity,
needed to relate the coherence time ($T_{0,2}$) to the coherence length ($r_0$), and $D_\tau(\Delta t)$ to $D_\tau(\Delta\vec x)$.
For a wind velocity of $v_w$ in azimuth $\phi_w$,
\begin{eqnarray}
A = &\sqrt{(r\cos\phi-r'\cos\phi')^2+(r\sin\phi-r'\sin\phi')^2} \nonumber \\
B= &\sqrt{(r\cos\phi+v_w\Delta t\cos\phi_w-r'\cos\phi')^2+(r\sin\phi+v_w\Delta t\sin\phi_w-r'\sin\phi')^2} \nonumber
\end{eqnarray}
A grid of model wind velocities was used to test the agreement between theoretical and measured angular structure functions.
In general, the agreement for a single turbulent flow velocity was poor --- the slope change in the model was sharper than in
the data.  Therefore, models with multiple layers (bulk wind velocities) were used.  The turbulence fields in different
layers were assumed to be independent, so that their contributions to the angular and delay structure functions
could be added.  Figures~\ref{fig3} and~\ref{fig4} show the agreement between data and model for single
and double layer models, for a scan on 1999, Night 109.   Adding even more layers 
to our models would have resulted in slightly better matches to the shapes of the measured structure
functions than in Figure~4.  However, it would not have changed the overall scaling mismatch.

There was not enough information in the angular structure functions to closely constrain the velocity of each component, but
the general shape of the overall velocity distribution was determined.

The overall amplitude scale for the FSM angles is uncertain by at least 20\%, enough to account for the discrepancy between
data and model seen in Figures~\ref{fig3} and~\ref{fig4} 
(note that the structure function is proportional to the {\it square} of the angles).
In addition, the model values reflect the average of the conditions at the two siderostats.  

\section {Results} \label{results}

\subsection {Delay Data}

Delay data (which passed the selection criteria described above) were obtained for 64 nights in 1999.  Table~1 gives a summary
of the data volume:  number of scans and total time span for each night, along with the mean values of the slope ($\beta$) and
coherence time ($T_{0,2}$).  For long scans, each 3~minute segment has been counted in the total in Table~\ref{tbl-1}.

\begin{figure}
\figurenum{5}
\plotone{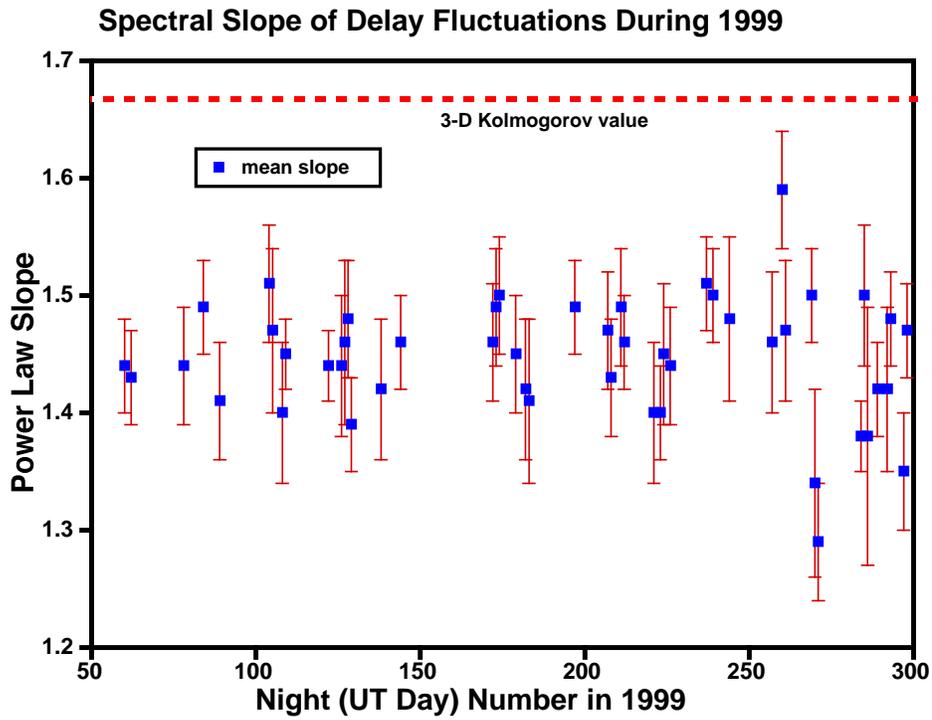}
\caption{Mean value of spectral slope ($\beta$) for each night in 1999 with 10 or more scans.  The error
bars represent the $1\sigma$ scatter about the mean value for that night. \label{fig5}}
\end{figure}

Figure~\ref{fig5} shows the mean fitted spectral slopes for the nights with $\ge 10$ usable scans.  The vertical bars represent
the $1\sigma$ scatter about the mean for that night.  The three dimensional Kolmogorov value of 5/3 is shown for comparison.

\begin{figure}
\figurenum{6}
\plotone{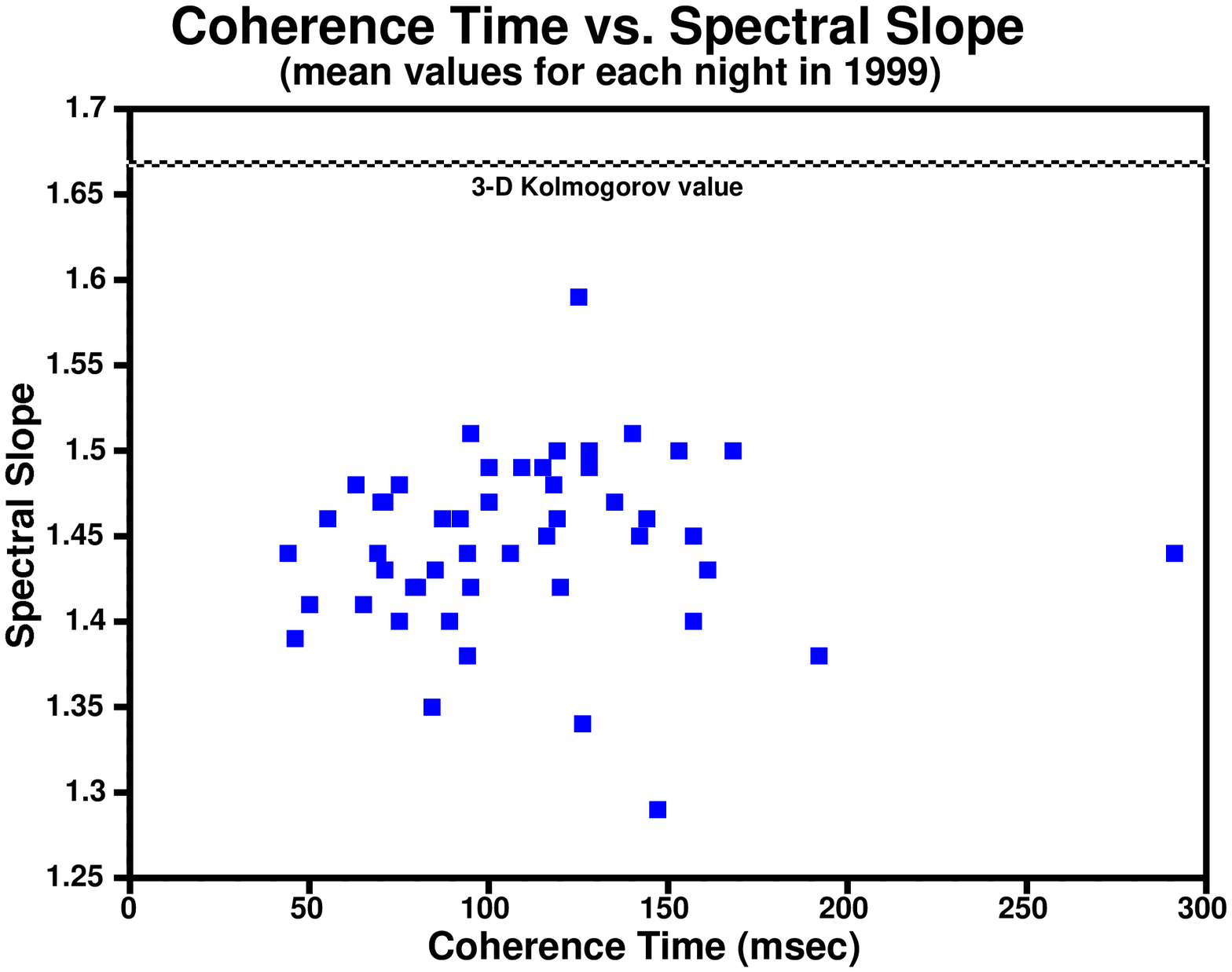}
\caption{Scatter plot of mean spectral slope against mean coherence time for each night in 1999 with 10 or
more scans. \label{fig6}}
\end{figure}

\begin{figure}
\figurenum{7}
\plotone{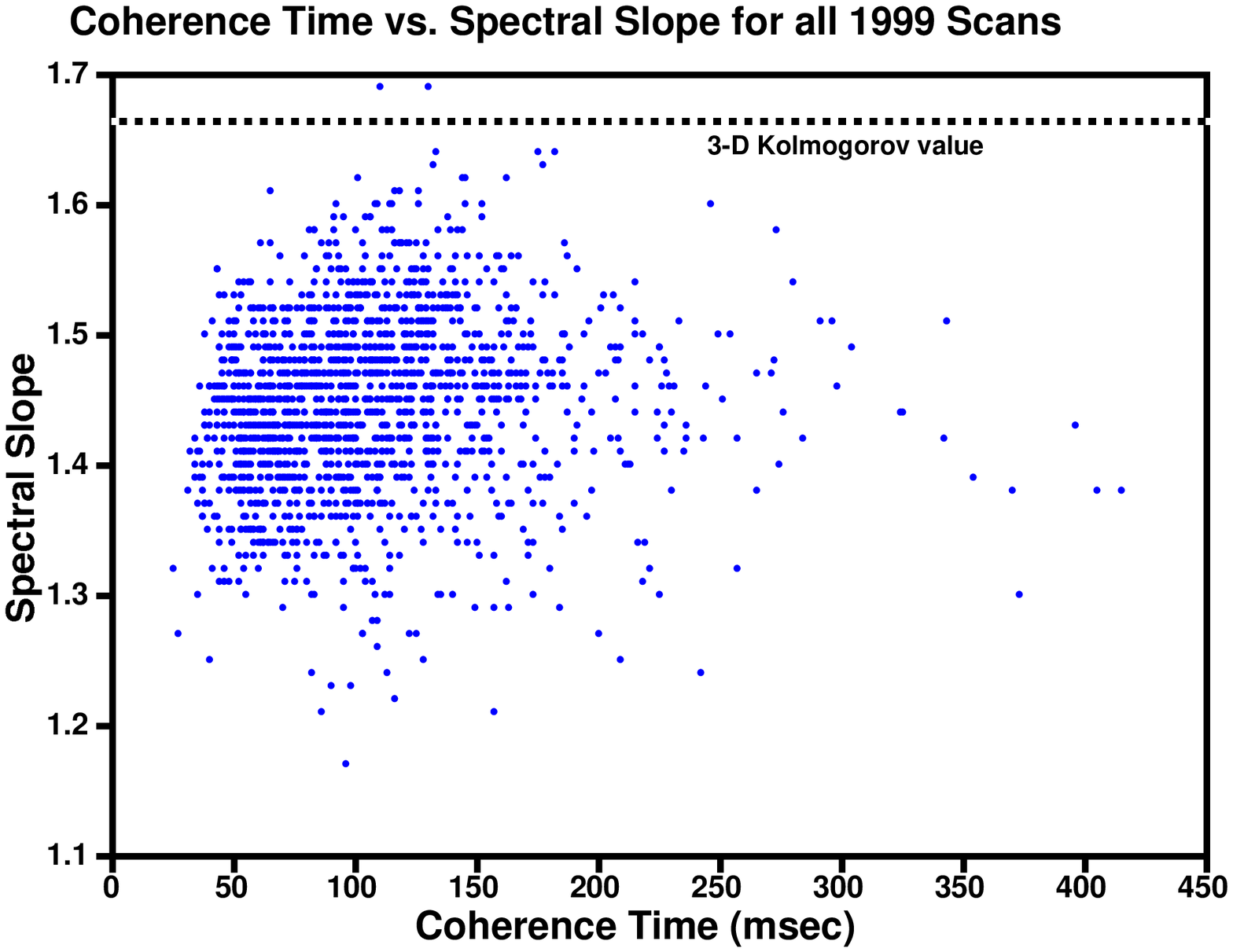}
\caption{Scatter plot of mean spectral slope against mean coherence time for each scan in 1999.  The
apparent quatization is due to the precision of 0.01 used in saving values of the fitted slope. \label{fig7} }
\end{figure}

Figure~\ref{fig6} is a scatter plot of the mean slope and coherence time for each night with $\ge 10$ usable scans.  
Figure~\ref{fig7} is a similar
plot for all the data, with one point per scan.
There appears to be no obvious correlation between the slope and coherence time.  The lower cutoff at $T_{0,2}\approx 30$~msec
is a selection effect:  for shorter coherence times, there were too many losses of lock to meet our selection criteria
(or the atmosphere was too noisy for the interferometer to operate at all).

\begin{figure}
\figurenum{8}
\plotone{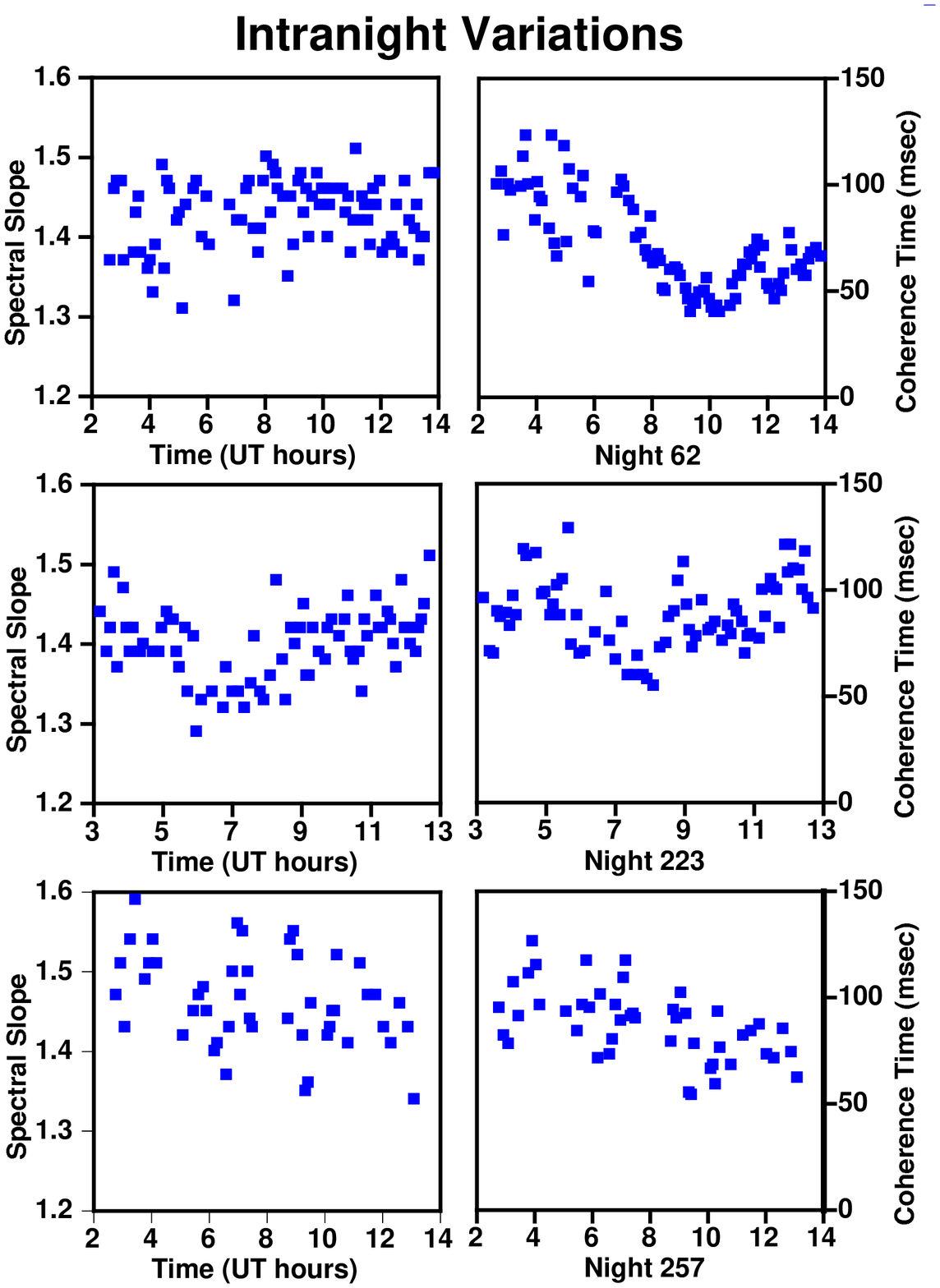}
\caption{Time history of spectral slope and coherence time during 1999 Nights 62, 223, and 257. \label{fig8}}
\end{figure}

The variations in $\beta$ and $T_{0,2}$ within individual nights did not fit into any obvious pattern.  
Figure~\ref{fig8} shows the
variations for nights 62, 223, and 257 in 1999 (this includes the two nights with the largest number of scans, and the
two nights with the largest time spans).  For most nights, there was no obvious trend in $\beta$ with time.  
The coherence time varied by factors of~2--4, sometimes on
timescales of $<1$~hr ({\it e.g.} on Night~62).  This result is consistent with reports of variations in seeing
on similar time scales ({\it e.g.} Martin et al. 1998).

An exponent of $\beta$ in the delay structure function corresponds to an exponent of $-(1+\beta)$ in the one-dimensional
spectrum of delay fluctuations (Armstrong \& Sramek 1982). For a Kolmogorov spectrum, the power spectrum exponent is
$-8/3$.

\subsection {Angle Tracking Data} \label{angles}

There were only four nights with extensive data recorded from both North and South star trackers.  Table~\ref{tbl-2} 
gives the results
of the velocity fitting for those nights.  
The time variations in the weights of the fitted wind velocities were small ($\sim 10$\%) within each night.

The weights in Table~2 represent the relative contributions to the coherence time ({\it i.e.} delay variations).  To get the
contribution of component $i$ to the coherence length, these weights should be scaled by $v_i^{-\beta}$, where $v_i$ is the
velocity of component $i$ and $\beta$ is the measured slope of the delay structure function (see Table~1).  Therefore, the
contributions of low velocity ($\sim 1\ {\rm m\ s^{-1}}$) turbulent flow to $r_0$ are more dominant than suggested by the weights in
Table~2.

\subsection {Long Time Intervals/Outer Scale Lengths} \label{outerscale}

If we assume that delay variations are due to a turbulence field convected past the telescopes (frozen flow), variations
in $T_{0,2}$ could be due either to variations in the flow velocity or to variations in the properties of the turbulence field.
Variations in the measured flow velocity (described above) within individual nights were small.  Therefore, variations
in $T_{0,2}$ must primarily reflect variations in the turbulence field.
As a measure of the total fluctuation amplitude, we used $D_\tau(50\ {\rm s})$: the delay structure function at
a time interval of 50~s.  This interval is approximately half the length of our shortest scans, and is therefore approximately
the longest interval over which we can get good statistics for $D_\tau$ in one scan.  By an interval of 50~s, $D_\tau$ had nearly
leveled off for most scans.  

\begin{figure}
\figurenum{9}
\plotone{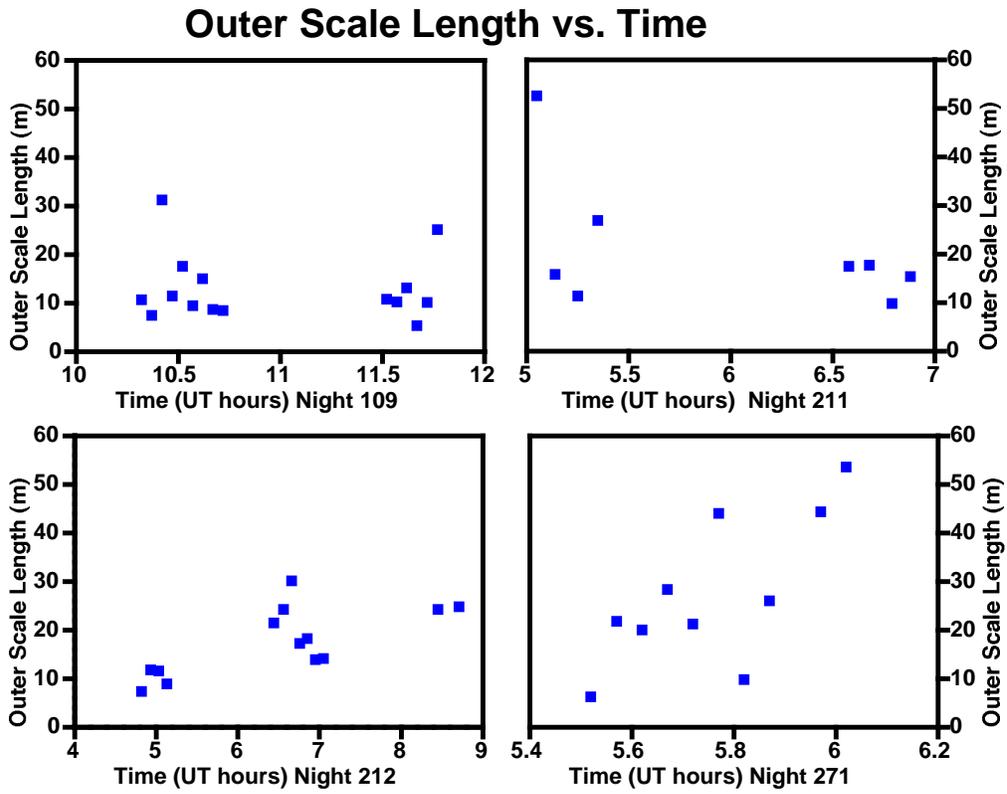}
\caption{Outer scale length vs. time for the four nights with star tracker data from both siderostats.  \label{fig9}}
\end{figure}

For those scans (on nights 109, 211, 212, and 271) with star tracker data, we can use $D_\tau(50\ {\rm s})$ to solve for an
outer scale size of the turbulence.
For $D_\tau(\Delta t)=c_0(\Delta t)^\beta$ and a set of velocities $v_i$ with weights (fractional
contribution to $D_\tau$) $f_i$, the structure function will saturate at a value:
\begin{equation}
D_\tau ({\rm max})=c_0 \sum_i\left[f_i\left({L_0\over v_i}\right)^\beta\right]
\end{equation}
Here $L_0$ is the structure function definition of the outer scale length: the spatial delay structure function 
$D_\tau(d)$ reaches a maximum value of $c_0L_0^\beta$.  
The derived outer scale values, for each scan with dual (north and south) angle tracking data, are shown in Figure~\ref{fig9}.

The decision to record dual star tracker data on these four nights was not based on the level of measured turbulence.
We therefore expect the results shown in Figure~9 to be representative of the conditions during the full set of 64 nights
listed in Table~1.  However, these results may not apply to the nights where the coherence time was too short for us to extract
atmospheric parameters.  We have no constraints on the outer scale length for those `noisy' nights.

An outer scale length in the refractivity power spectral density is generally represented with the von Karman model
(Ishimaru 1978):
\begin{equation}
\Phi_N(\kappa)\propto\left[\kappa^2+\left({1\over {\cal L}_0}\right)^2\right]^{-11/6}
\end{equation}
The spatial frequency $\kappa=2\pi/\lambda$, with $\lambda$ the wavelength of the fluctuation.  For the generalization to
non-Kolmogorov slopes, the exponent is $-(1+\beta/2)$.

In order to obtain the correspondence between the power spectral density outer scale ${\cal L}_0 ({\rm PSD})$ and the
structure function outer scale $L_0 (D_\tau)$, equation(14) was first transformed to a refractivity structure function.  This
refractivity structure function was then numerically integrated to yield a delay structure function.  The results on the
outer scale correspondence, for three representative values of $\beta$, are:
\begin{eqnarray}
L_0 (D_\tau)=&0.94 {\cal L}_0 ({\rm PSD})\quad(\beta=5/3) \nonumber \\
=&1.03 {\cal L}_0 ({\rm PSD})\quad(\beta=1.50) \\
=&1.18 {\cal L}_0 ({\rm PSD})\quad(\beta=1.35) \nonumber
\end{eqnarray}

\section {Discussion}

\subsection {Spectral Slope}

Our measured power law slopes for short time scale delay variations were largely in the range 1.40--1.50, and were in all
cases shallower than the three-dimensional Kolmogorov value of 5/3.  The spectral slope was not correlated with the coherence
time, at least when atmospheric conditions were stable enough for operation of the interferometer.

The Kolmogorov spectrum is based on dimensional considerations \citep{tat61}. Measurements 
of strong turbulence in a variety of fluids have shown good agreement to Kolmogorov spectra \citep{gra62, fri90}.  However,
atmospheric conditions during astronomical observations involve much weaker turbulence.  Intermittent turbulence
\citep{fri78} may decrease the slope of the spectrum under these conditions.

\citet{bus95} analyzed atmospheric fluctuations with interferometric measurements from Mt. Wilson, 
on baselines from 3~to 31~m in length.
Based on power spectral densities, they found a mean slope slightly shallower than Kolmogorov, by 0.12 (equivalent to
a slope of 1.55 for delay structure functions).  \citet{bes92} 
also used data from Mt. Wilson.  However, all their short time scale measurements were made with a laser
distance interferometer, over a horizontal path of length $\sim 10$~m, located 3~m above the ground.  They measured
spectral slopes shallower than the Kolmogorov value by nearly 0.30 ({\it i.e.} a structure function exponent of 
$\beta\approx 1.40$).

Our results agree with those of \citet{bes92}, although our data were taken on approximately vertical line-of-sight
paths through the entire atmosphere, and theirs on much shorter paths near the ground.  Our results disagree with those of
\citet{bus95}, if we use the standard  correspondence between the exponents of the
power spectral density ($\gamma$) and structure function ($\beta$):  $-\beta-1\leftrightarrow\gamma$ \citep{arm82}.  However, 
\citet{bes92} report the surprising result that their value of $\gamma$ is often 0.1 or 0.2 steeper than
$-\beta-1$ for the same data (their actual comparison is between power spectral density and Allan variance, whose slope is 
nearly equal to that of the structure function at short time scales).  They speculate that the discrepancy may result
from occasional bursts with $\beta>2$, for which the $-\beta-1\leftrightarrow\gamma$ connection does not hold.

For $D_\tau\propto(\Delta t)^\beta$, the dependence of coherence length $r_0$ upon wavelength $\lambda$ will be\hfil\break 
$r_0\propto\lambda^{2/\beta}$, and the seeing $(\theta)$ will vary as $\theta\propto\lambda^{1-2/\beta}$.  A value of
$\beta=5/3$ gives $\theta\propto\lambda^{-0.2}$, while $\beta=1.40$ gives $\theta\propto\lambda^{-0.43}$.  Numerous reports
of exceptional seeing at infrared wavelengths ({\it e.g.} diffraction rings in images with the Palomar 5~m telescope
at wavelengths of $10\ \mu{\rm m}$ and even $5\ \mu{\rm m}$) are most easily explained with a steep dependence of seeing
{\it vs.} wavelength.

For astrometry with ground-based interferometry, a low value of $\beta$ has favorable consequences.  Over a baseline of
length $B$, the instantaneous delay uncertainty $\Delta\tau_{\rm atm}$ of the atmosphere will be:
\begin{displaymath}
\Delta\tau_{\rm atm}\sim {\lambda\over 2\pi}\left({B\over r_0}\right)^{\beta/2}
\end{displaymath}
Here $r_0$ is the coherence length at wavelength $\lambda$, and an infinite outer scale length is assumed.  For a finite
outer scale length $L_0$, the baseline length $B$ should be replaced with $L_0$.

\subsection {Turbulence speed and height}

Our derived wind velocity for the turbulent layer(s) (section~\ref{angles}) was low:  1--4$\ {\rm m\ s^{-1}}$ 
(plus a small $10\ {\rm m\ s^{-1}}$ component on one
night).  For interferometric measurements on only one star at a time, we have no direct constraint on the height of the 
turbulence.  However, the very low velocities suggest that the turbulence was at low altitudes, perhaps even within 100~m
of the surface.  \citet{tre95} derived a low altitude ($<45$~m) for the majority of the turbulence seen at the
Mt. Wilson Infrared Spatial Interferometer, based on a correlation between fluctuations seen with the starlight
interferometer and a laser distance interferometer.

Simultaneous observations of two (or more) stars with the same pair of siderostats would allow a direct determination of 
turbulent height.  

\subsection {Outer Scale Lengths}

Our measured outer scale lengths are mostly in the range 10--25~m, in agreement with results reported by others
\citep{cou88, zia94}.  All our values are less than half the 110~m length
of our baseline, giving us confidence that our results are not significantly corrupted by the finite length of the
baseline.  Because these outer scale values are based on simultaneous measured delay {\it and} angle time series 
on a long baseline,
they are less sensitive to modeling assumptions than most previously published results.  \citet{sha92} derived
the effect of a finite outer scale size on the accuracy of narrow angle astrometry.  
For $L_0<<B$ ($B$ is the baseline length), the accuracy
improvement over an infinite outer scale, for $\beta=5/3$, is $(L_0/B)^{1/3}$.  
For $L_0\approx 15$~m and $B=110$~m, this factor is
$\approx 0.5$.  For longer baselines, the accuracy improves as $B^{-1}$, rather than as $B^{-2/3}$ for an infinite outer
scale.

\acknowledgments

This work was performed at the Infrared Processing and Analysis Center, California Institute of Technology, 
and at the Jet Propulsion Laboratory, California Institute of Technology, under contract with the National Aeronautics
and Space Administration.
Data were obtained at the Palomar Observatory using the NASA Palomar Testbed Interferometer, which is supported by
NASA contracts to the Jet Propulsion Laboratory.  Science operations with PTI are possible through the efforts of the
PTI Collaboration (\url{http://huey.jpl.nasa.gov/palomar/ptimembers.html}).

\begin{deluxetable}{rrrcr}
\tablewidth{0pt}
\tabletypesize{\scriptsize}
\tablecaption{Summary of Atmospheric Data From 1999.  \label{tbl-1}}
\tablehead{
\colhead{Night} & \colhead{No. Scans} & \colhead{Time Span} & \colhead{Mean~$\beta$} & \colhead{Mean~$T_{0,2}$}
}
\startdata
59&6&3.5 hr&1.44&117 msec \\
60&67&9.8 hr&1.44&69 msec \\
62&87&11.3 hr&1.43&71 msec \\
78&48&9.7 hr&1.44&106 msec \\
84&17&2.9 hr&1.49&100 msec \\
89&22&6.5 hr&1.41&50 msec \\
101&7&2.9 hr&1.34&53 msec \\
104&26&8.6 hr&1.51&95 msec \\
105&32&9.5 hr&1.47&70 msec \\
106&7&0.8 hr&1.45&58 msec \\
108&22&5.4 hr&1.40&157 msec \\
109&32&5.9 hr&1.45&116 msec \\
122&12&1.4 hr&1.44&44 msec \\
126&21&3.4 hr&1.44&291 msec \\
127&43&5.3 hr&1.46&144 msec \\
128&34&6.8 hr&1.48&75 msec \\
129&32&7.1 hr&1.39&46 msec \\
138&18&5.3 hr&1.42&120 msec \\
141&6&1.3 hr&1.44&52 msec \\
144&12&4.9 hr&1.46&55 msec \\
145&4&1.0 hr&1.44&40 msec \\
172&24&7.8 hr&1.46&92 msec \\
173&47&4.4 hr&1.49&128 msec \\
174&41&4.5 hr&1.50&168 msec \\
175&9&2.8 hr&1.39&130 msec \\
177&8&3.2 hr&1.37&56 msec \\
178&16&7.8 hr&1.43&85 msec \\
179&32&7.9 hr&1.45&142 msec \\
180&9&2.2 hr&1.45&227 msec \\
181&7&0.7 hr&1.40&94 msec \\
182&12&5.7 hr&1.42&95 msec \\
183&11&2.7 hr&1.41&65 msec \\
197&15&4.0 hr&1.49&115 msec \\
205&5&1.3 hr&1.51&93 msec \\
207&29&7.4 hr&1.47&135 msec \\
208&46&8.3 hr&1.43&161 msec \\
211&41&7.2 hr&1.49&109 msec \\
212&23&4.8 hr&1.46&119 msec \\
214&4&0.5 hr&1.41&85 msec \\
221&31&7.1 hr&1.40&75 msec \\
223&74&9.5 hr&1.40&89 msec \\
224&19&5.0 hr&1.45&157 msec \\
226&34&5.3 hr&1.44&94 msec \\
231&9&7.5 hr&1.50&151 msec \\
237&23&5.4 hr&1.51&140 msec \\
239&27&8.2 hr&1.50&119 msec \\
244&21&4.9 hr&1.48&118 msec \\
257&47&10.3 hr&1.46&87 msec \\
258&6&6.2 hr&1.44&134 msec \\
260&25&5.1 hr&1.59&125 msec \\
261&36&9.3 hr&1.47&71 msec \\
269&40&6.8 hr&1.50&128 msec \\
270&43&5.2 hr&1.34&126 msec \\
271&16&3.1 hr&1.29&147 msec \\
284&13&7.5 hr&1.38&94 msec \\
285&13&8.0 hr&1.50&153 msec \\
286&13&5.8 hr&1.38&192 msec \\
289&23&8.3 hr&1.42&79 msec \\
292&28&7.9 hr&1.42&80 msec \\
293&32&10.1 hr&1.48&63 msec \\
294&5&3.5 hr&1.52&92 msec \\
297&17&5.5 hr&1.35&84 msec \\
298&19&5.0 hr&1.47&100 msec \\
\enddata

\tablecomments{The five columns are:  1) UT Day number corresponding to a night of observations, 2) Number of usable scans
for that night, 3) Time spanned by the scans on that night, 4) Mean spectral slope $(\beta)$ of the scans on that night,
and 5) Mean coherence time $(T_{0,2})$ for the scans on that night.}
\end{deluxetable}
\clearpage
 
\begin{deluxetable}{rrrccccc}
\tabletypesize{\scriptsize}
\tablecaption{Best Fit Turbulent Flow Velocities.  \label{tbl-2}}
\tablewidth{0pt}
\tablehead{
\colhead{Night} & \colhead{Time Span} & \colhead{Vel. 1} & \colhead{Weight (N)} & \colhead{Weight (S)} &
\colhead{Vel. 2} & \colhead{Weight (N)} &\colhead{Weight (S)}
}
\startdata
109&1.5 hr&$1\ {\rm m\ s^{-1}}$&0.34&0.42&$4\ {\rm m\ s^{-1}}$&0.66&0.58 \\
211&1.8 hr&$0.8\ {\rm m\ s^{-1}}$&0.14&0.16&$3\ {\rm m\ s^{-1}}$&0.86&0.84 \\
212\tablenotemark{a}&3.9 hr&$0.5\ {\rm m\ s^{-1}}$&0.09&0.10&$2\ {\rm m\ s^{-1}}$&0.63&0.80 \\
271&0.4 hr&$1\ {\rm m\ s^{-1}}$&0.50&0.57&$4\ {\rm m\ s^{-1}}$&0.50&0.43 \\
\enddata

\tablenotetext{a}{A third component, with $10\ {\rm m\ s^{-1}}$ velocity, and weight 0.28 (N) and 0.10 (S) was 
needed to fit the star tracker data for Night 212.}

\tablecomments{The eight columns are:  1) UT Day number corresponding to a night of observations, 2) Time spanned by the
scans with star tracker data on that night, 3) Velocity of component one, 4) Weight of component one for the North siderostat,
5) Weight of component one for the South siderostat, 6) Velocity of component two, 
7) Weight of component two for the North siderostat,
8) Weight of component two for the South siderostat.}

\end{deluxetable}

\end{document}